\documentclass[journal]{IEEEtran}
\usepackage{cite}
\usepackage{hyperref, color}
\definecolor{linkColor}{rgb}{1,0,0}
\hypersetup{pdfborder={0 0 0},colorlinks=true,urlcolor=linkColor,citecolor=linkColor}

\ifCLASSINFOpdf
   \usepackage[pdftex]{graphicx}
\else
\fi
\usepackage{amsmath}
\usepackage{algpseudocode} 
\usepackage{algorithm}

\begin{document}
%
\title{Distributed Reconstruction Algorithm for Electron Tomography with Multiple-scattering Samples}
%
%
%

\author{David~Ren,
        Michael~L.~Whittaker,
        Colin~Ophus,
        and~Laura~Waller
\thanks{D. Ren and L. Waller are with the Department
of Electrical Engineering and Computer Sciences, University of California, Berkeley,
CA, 94720 USA e-mail: david.ren@berkeley.edu; waller@berkeley.edu.}%
\thanks{M. Whittaker is with the Energy Geosciences Division, Lawrence Berkeley National Laboratory, CA, 94720.}%
\thanks{C. Ophus is with the National Center for Electron Microscopy, Molecular Foundry, Lawrence Berkeley National Laboratory, CA, 94720.}%
\thanks{Manuscript received April 19, 2005; revised August 26, 2015.}}

%
%

\markboth{Journal of \LaTeX\ Class Files,~Vol.~14, No.~8, August~2015}%
{Shell \MakeLowercase{\textit{et al.}}: Bare Demo of IEEEtran.cls for IEEE Journals}
%



\maketitle

\begin{abstract}
Three-dimensional electron tomography is used to understand the structure and properties of samples in chemistry, materials science, geoscience, and biology. With the recent development of high-resolution detectors and algorithms that can account for multiple-scattering events, thicker samples can be examined at finer resolution, resulting in larger reconstruction volumes than previously possible. In this work, we propose a distributed computing framework that reconstructs large volumes by decomposing a projected tilt-series into smaller datasets such that sub-volumes can be simultaneously reconstructed on separate compute nodes using a cluster. We demonstrate our method by reconstructing a multiple-scattering layered clay (montmorillonite) sample at high resolution from a large field-of-view tilt-series phase contrast transmission electron microscopty dataset.
\end{abstract}


%
\IEEEpeerreviewmaketitle

\section{Introduction}
Electron tomography (ET) enables an unprecedented view of microscopic samples, where 3D reconstructions of the local structure have led to scientific discoveries in biology, materials science and chemistry structure~\cite{Midgley:2009electron,Leary:2012recent,Bharat:2015advances}. In a bright field transmission electron microscopy (TEM) experiment, a plane wave illuminates the sample and phase delays in the electron wave are induced by the sample's electrostatic potential. Phase contrast can be obtained by slightly defocusing the image of the sample, and the image contrast will be linear with respect to the cumulative phase as long as the sample is weakly-scattering. Many biological samples and thin foils meet this criteria; as such, classical tomography methods that rely on the projection slice theorem\cite{Gilbert:1972iterative,Muller:2015theory,Pryor:2017genfire} can be directly applied to solve for 3D structures when the weak phase approximation holds. However, to ensure the validity of the weakly-scattering assumption, the sample should be thin, and this poses a great challenge to sample preparation. For thicker samples or materials with larger atomic numbers, nonlinear scattering effects become non-negligible.

In addition, many samples cannot tolerate high electron dose -- these samples are often vitrified in medium at cryogenic temperature to reduce damage from the electron beam, a technique known as cryo-EM for 2D imaging or cryo-ET for 3D imaging~\cite{Bharat:2015advances}. Electron beam damage is a complex process that is not fully understood~\cite{Egerton:2021radiation}, but is roughly inversely correlated with atomic number and occurs more rapidly at surfaces and defects. Therefore, most matter on Earth's surface is generally beam sensitive in an electron microscope, being composed of light elements that are frequently hydrated and imaged with cryoEM/cryoET~\cite{Hochella:2019natural,Whittaker:2020}. Being able to model the multiply-scattering events between the sample and the beam probe and more efficiently `use' the electrons counted in the images can reduce the amount of dose needed to achieve the same reconstructed SNR~\cite{Ren:2020}. 

To image thicker and more diverse samples, multiple scattering and the non-linearity in the image formation process should be taken into account~\cite{Van:2012method,Suzuki:2014high,Kamilov:2015learning,Ren:2020,Pelz:2021phase,Liu:2017seagle}. This can be done by incorporating such phenomena into the forward model, for example by implementing a multi-slice (beam propagation) model~\cite{Van:2012method,Kamilov:2015learning,Ren:2020}, which represents the 3D scattering process using a sequence of 2D layers of transmittance functions that cause beam absorption and phase delay. The input plane wave is propagated through the layers, assuming a fixed distance of free space in between. This method is intuitive, robust, and efficient to implement. Multi-slice methods are powerful, but more computationally expensive than traditional projection-based methods, and often need to be run dozens or hundreds of times inside an iterative reconstruction loop.

Computational requirements are further increased by recent advances in direct electron detector technology which have greatly improved imaging throughput, both in frame rate and pixel count. Many detectors have demonstrated capacity to capture images with up to $8k\times8k$ pixels~\cite{Hattne:2019microed,Nakane:2020single, Takaba:2021protein,Mcmullan:2014comparison}. In our study, for example, a Gatan K3 detector is used, and the images captured have $5760\times4092$ pixels. Given that 3D techniques generally capture dozens or hundreds of images in a tilt-series, the result is that very large volumes (in terms of voxels) can be reconstructed.

With these hardware improvements, more computational resources are required to process the increasing amount of tomography data. The scale of the datasets usually exceeds the capacity of a modern single-node computer; hence, parallelization is needed to decompose the reconstruction into multiple parallel sub-problems. Previous work in linear tomography has extensively used parallel computing~\cite{Fernandez:2008efficient, Gordon:2005component,Basu:2000n,Basu:2002n,Hidayetoglu:2020petascale,Hidayetouglu:2019memxct,Wang:2019consensus}, but requires linear projections or tomographic matrix sparsity. Since multiple scattering is a nonlinear interaction between the wave and the sample, a matrix cannot be written to represent the scattering process. Consequently, these methods are not compatible with existing distributed computing strategies.

Here, we propose a distributed algorithm that preserves the coherent diffraction effects in the tilt-series, and thus works for multiple-scattering forward models. We first show that by carefully cropping out particular regions of each raw image in the original tilt-series, any given sub-volume within the full volume can be reconstructed, though artifacts may occur due to contributions from outside of the sub-volume. Thus, we can define many sub-volumes within the full volume and reconstruct them all in parallel on separate compute nodes, while still accounting for multiple-scattering effects. The sub-volumes are then stitched together to form the full-volume reconstruction. We demonstrate the performance of the algorithm experimentally by solving small tomography problems while varying different parameters. Both reconstruction time and mean square error (MSE) are reported. Finally, we demonstrate the tomographic reconstruction of a large volume consisting of clay minerals vitrified in aqueous solution. To our knowledge, this is the largest volume (in number of voxels) ever reconstructed in cryo-ET, with a size of ($0.73(x)\times1.00(y)\times1.73(z) \mu \text{m}^3$) and resolution of 1.82 $\rm{\AA}^3$/voxel. At this resolution, not only do we see unprecedented microscopic features, but also we can visualize and understand macroscopic sample structure immersed in the solution. In all, our method offers the following advantages:
\begin{itemize}
    \item It is model independent as we only manipulate the tilt-series before the reconstruction. The choice of tomography algorithm is thus decoupled, ensuring compatibility with any multiple-scattering tomography model.
    \item It requires no inter-node communication -- during the reconstruction, all parallel compute nodes are completely independent from each other. Therefore, the reconstruction speed of one node does not impact the speed of others, and results in less overall process idle time. 
    \item It does not require one to possess deep knowledge of computer architecture, because implementing it does not need low-level architecture-specific optimization such as exploiting the matrix sparsity and other structural properties of the linear inverse problem.
\end{itemize}

\begin{figure*}[!thbp]
	\centering
	\includegraphics[width=\textwidth]{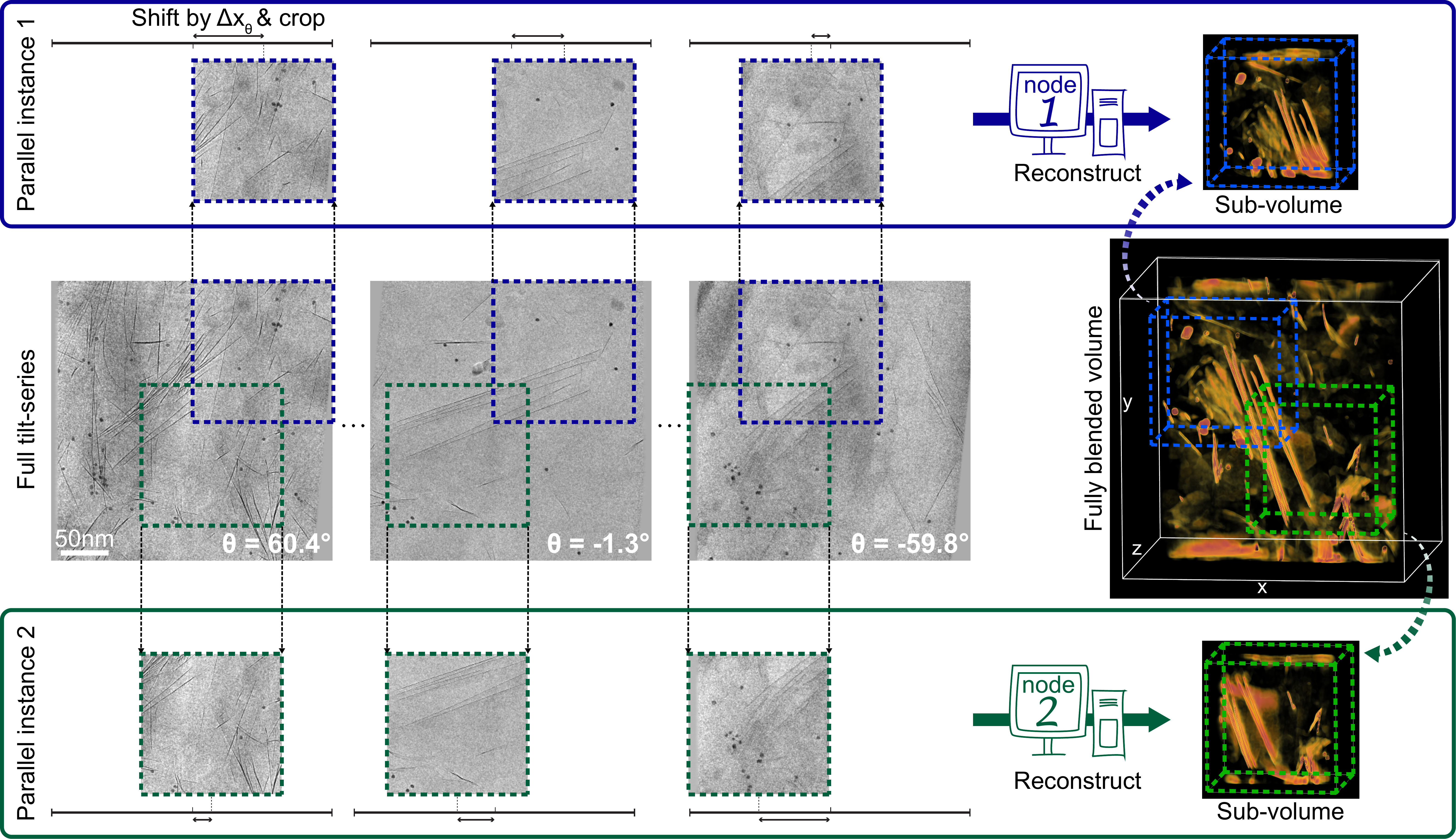}
	\caption{Conceptual illustration of our distributed reconstruction algorithm. Two examples of parallel nodes are shown in this figure, in blue and green, respectively. Each correspond to a different sub-volume of the sample, the tilt-series of which are shifted by $\Delta x_{\theta}=x_0\mathrm{cos}(\theta)+z_0\mathrm{sin}(\theta)$ and cropped from the full tilt-series dataset depending on the position of the sub-volume with respect to the full volume. After all sub-volumes are reconstructed, they are stitched together to form the large-volume reconstruction.}
	\label{Fig:schematics}
\end{figure*}

\section{Background}
\subsection{Image Formation}
In classical electron tomography, a sample is placed on a substrate and tilted to different angles with respect to the imaging system. At each tilt angle, the electron beam is illuminated onto the sample. After the interaction between the sample and the probe beam, an image is formed by the generalized linear operator $\mathcal{A}${}:
\begin{equation}
    I(x,y) = \mathcal{A}\{f(x,y,z)\} + e,
\end{equation}
\noindent where $I(x,y)$ is the measurement, $f(x,y,z)$ is the sample's electric potential, $\mathcal{A}$ is the forward model operator, and $e$ is the noise associated with the measurement process, modeled here as additive noise with a specified distribution. In Atomic Electron Tomography (AET)~\cite{Xu:2015three,Yang:2017deciphering}, the measured signal $I(x,y)$ is modelled as a linear axial projection of the sample's electric potential $f(x,y,z)$, and the linear operator $\mathcal{A}$ represents the additive projection matrix under discretization. In cryo-ET, contrast transfer function (CTF) corrections are usually performed on the defocused bright-field intensity measurements in order to obtain a signal that is then assumed to be a linear projection of the electric potential. Therefore, in this case, not only is $\mathcal{A}$  a linear projection matrix, but also incorporates another linear operator modelling the CTF.

With forward models that include multiple scattering (e.g. the multi-slice method), we can account for the nonlinear interaction between the probe beam and the sample~\cite{Van:2012method,Kamilov:2015learning,Chowdhury:2019,Ren:2020}. In general such models can be denoted by a generalized nonlinear scattering operator $\mathcal{S}$:
\begin{equation}
    I(x,y) = \left|\mathcal{S}\{f(x,y,z);E_i(x,y,z)\}\right|^2 + e.
\end{equation}
These forward models predict the measured intensity for an estimated sample, given an incident illumination beam $E_i(x,y,z)$. For example, in~\cite{Ren:2020}, the multi-slice scattering operator is a recursive method that models the propagation of the beam through a series of 2D projected potentials. The operator then outputs exit waves that are nonlinear with respect to the underlying sample. In the end, the intensity of the complex exit wave is calculated. For nonlinear scatttering operators, $\mathcal{S}$ cannot be written in the form of a matrix, and so matrix inversion techniques are not applicable.

\subsection{Inverse Model}
Given a forward model, we then set up an inverse problem to recover the 3D sample from a 2D tilt-series. Iterative solvers are popular among both linear and nonlinear tomography problems, offering flexible choice of forward model and straightforward incorporation of constraints such as sample support region or prior knowledge. The inverse problem can be formulated as an optimization problem:
\begin{equation}
    f^* = \operatorname*{arg\,min}_{f}\{\mathcal{D}(f;I) + \beta\mathcal{R}(f)\},
    \label{eq:inverse_model}
\end{equation}
\noindent where $\mathcal{D}(\cdot)$ is the data fidelity term, which measures the distance between predicted measurements and the true measurements. The $l_2$ norm is a commonly chosen metric for measuring distance, along with other variants~\cite{Yeh:2015experimental,Kamilov:2015learning}.  $\mathcal{R}(\cdot)$ is an optional regularization term to incorporate constraints, and is especially important when solving an under-determined problem~\cite{Ren:2020,Pham:2020resire,Pham:2020adaptive}. The scalar term $\beta$ is used to balance the strength of the regularization enforcement and the data fidelity term. After choosing a value for $\beta$, the solution $f^*$ is that which minimizes Eq.~\eqref{eq:inverse_model}.

To minimize Eq.~\eqref{eq:inverse_model}, one can use convex optimization techniques.  For instance, GENFIRE~\cite{Pryor:2017genfire} uses well-known projection-based methods, where projections and back-projections are performed simultaneously with applications of constraints in different domains. Gradient-based algorithms have also been proposed~\cite{Gilbert:1972iterative, Van:2012method, Ren:2020}, where gradients with respect to the data fidelity term as well as the regularization term are calculated and the solution is refined iteratively. Prior knowledge of the system and the solution can be incorporated via proximal methods~\cite{Parikh:2014proximal}.



\section{Methods}
\subsection{Reconstructing a sub-volume}

In this section, we will show that a subset of the full volume can be reconstructed independently by shifting and cropping each raw image in the tilt-series appropriately. Each sub-volume within the full 3D sample volume can then be reconstructed in a distributed fashion, as shown in Fig.~\ref{Fig:schematics}. To calculate which parts of each raw image to crop for a given sub-volume, we use the mathematics of linear projection, also known as the Radon transform~\cite{radon20051,radon:1986determination}. We first look at the 2D projection image formulation for single-axis tilt along the $y$-axis. The 2D projected image $I(x,y)$ at tilt angle $\theta$ is related to the volume-of-interest through a Radon transform:
\begin{align}
  &\begin{aligned}
    I(x,y;\theta) = \iint f&(x',y,z') \times\\
      &\delta(x'\mathrm{cos}(\theta)+z'\mathrm{sin}(\theta)-x)\,\mathrm{d}x'\mathrm{d}z',
  \end{aligned}
\end{align}
where $f(x,y,z)$ is the 3D sample and $\delta(\cdot)$ denotes Dirac delta function. A full projection dataset is formed by rotating the 3D sample to different angles $\theta$. The tilt axis is assumed to be at the \emph{center} of the volume. However, when a sub-volume of the 3D sample is considered, the center of it does not necessarily coincide with the center of the full 3D volume, and further manipulation is required to relate the tilt-series of the sub-volume to that of the full volume.

Consider the tilt axis of a sub-volume. It is parallel with the true tilt axis of the full volume during the experiment, and translated along $x$ and $z$ by $x_0$ and $z_0$, respectively. This is equivalent to the sample being shifted in opposite directions. The new projected images $I'(x,y;\theta)$ are then:
\begin{align}
  &\begin{aligned}
    I'(x,y;\theta) = \iint f&(x'-x_0,y,z'-z_0) \times\\
      &\delta(x'\mathrm{cos}(\theta)+z'\mathrm{sin}(\theta)-x)\,\mathrm{d}x'\mathrm{d}z'.
  \end{aligned}  
\end{align}
After performing a change of variables ($x''=x'-x_0$ and $z''=z'-z_0$) and simplification, the relationship becomes

\begin{align}
  &\begin{aligned}
    I'(x,y;\theta) = \iint f&(x'',y,z'') \times\\
      &\delta [x''\mathrm{cos}(\theta)+z''\mathrm{sin}(\theta)-\\
      &(x-x_0\mathrm{cos}(\theta)-z_0\mathrm{sin}(\theta))]\,\mathrm{d}x''\mathrm{d}z''.
  \end{aligned}  
\end{align}
The new projections can now be related to the original projection as:
\begin{equation}
    I'(x,y;\theta) = I(x-x_0\mathrm{cos}(\theta)-z_0\mathrm{sin}(\theta),y;\theta).
    \label{eq:shfit_tilt}
\end{equation}
Given a tilt angle $\theta$, the new projection corresponding to the tilt axis of the sub-volume is simply a shift along the $x$-axis with an amount of $\Delta x_{\theta} = x_0\mathrm{cos}(\theta)+z_0\mathrm{sin}(\theta)$. With the new tilt-series $I'(x,y;\theta)$, the set can be cropped while ensuring that the subset is centered around the tilt axis of the sub-volume. Because the result suggests a global shift of the projection along the $x$-axis for each tilt angle $\theta$, it preserves the multiple scattering and diffraction effects in the original tilt-series. As such, new sets of tilt-series corresponding to different sub-volumes can be calculated independently from the original tilt-series. After that, each new set of tilt-series can be used to reconstruct the sub-volumes simultaneously on different compute nodes, in order to improve compute speed.

\subsection{Sub-volume Overlap}
\label{subsec:overlap}

When splitting a full volume into smaller sub-volumes, overlap between adjacent sub-volumes is necessary both to avoid empty areas in the volume, as well as to avoid artifacts due to diffraction and multiple scattering, where electrons or light scatters outside the edges of the sub-volume~\cite{Chowdhury:2019}.

We can derive the minimum amount of overlap in order to avoid empty areas and ensure coverage of the full 3D volume. From a tilt series of size $N_x \times N_y$ pixels, the size of the reconstructed volume is $N_x \times N_y \times N_z$ voxels, where each is the number of pixels along each of the three axes, respectively. Since the sample is rotating with respect to the $y$-axis, the support of the reconstructed sample within the volume is an inscribed elliptic cylinder with semimajor axis $N_x/2$, semiminor axis $N_z/2$ and height $N_y$. If $N_x = N_z$, the support becomes a cylinder with radius $N_x/2$. When we concatenate the sub-volume cylinders together to form a 3D volume, there will still be areas in the volume that remain empty (i.e. the four corners) unless there is sufficient overlap between the cylinders.

\begin{figure}[!htbp]
	\centering
	\includegraphics[width=0.9\linewidth]{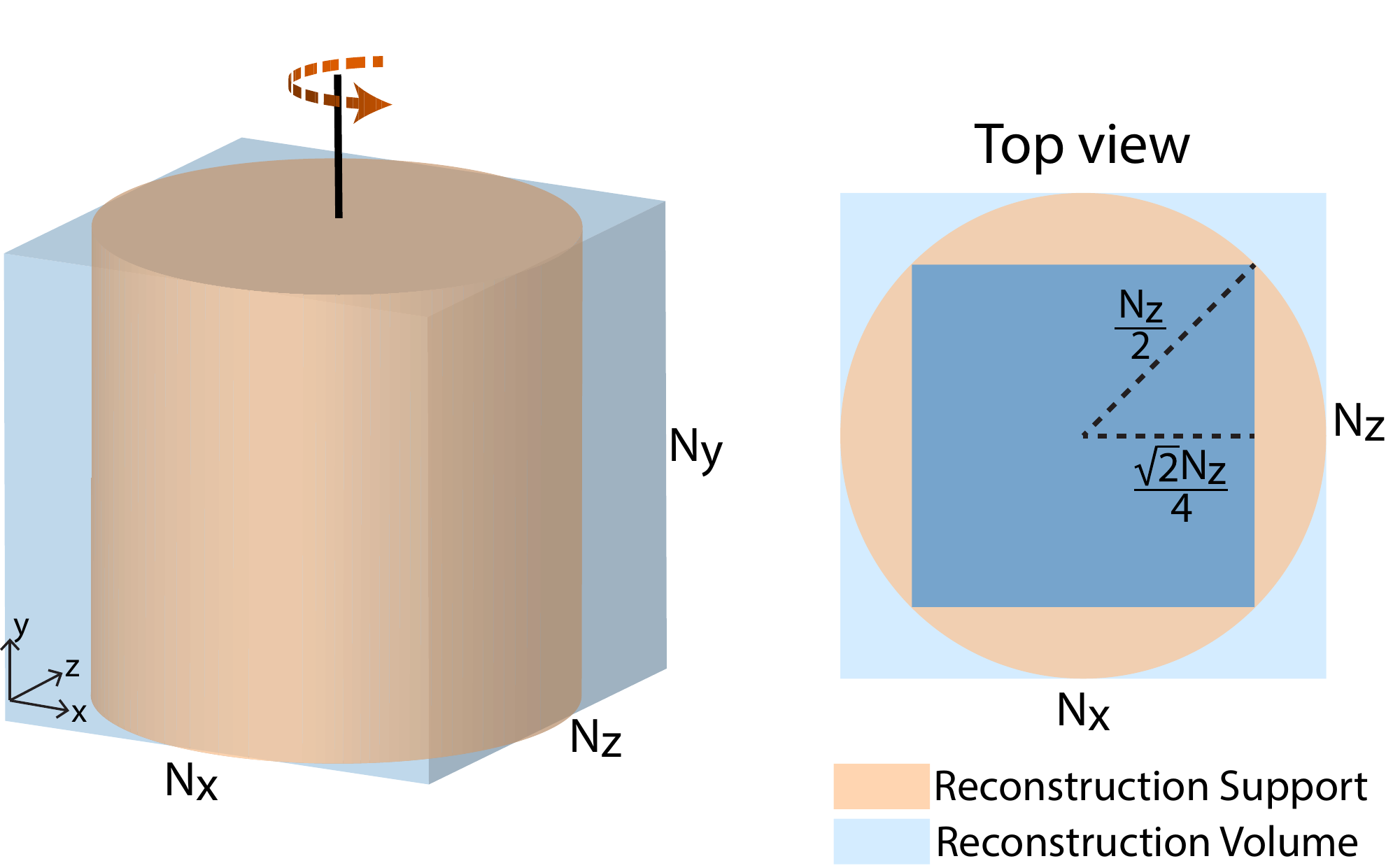}
	\caption{Illustration for calculating the minimum overlap between sub-volumes that covers the entire volume, assuming $N_x=N_z$. The reconstructed volume has cylindrical support and the inscribed blue square with side length of $\sqrt{2}/2N_z$ is used for final volume stitching. As such, the minimum overlap required is approximately 30\%. }
	\label{Fig:overlap_schematics}
\end{figure}

In our method, we overlap the adjacent sub-volumes, and only consider the cuboid inscribed within the cylinder ( Fig.~\ref{Fig:overlap_schematics}) for each sub-volume reconstruction. By geometry, the inscribed cuboid has length and width of $\sqrt{2}/2N_x$ and $\sqrt{2}/2N_z$, respectively. This corresponds to a minimum required overlap of 30\% along both $x$-axis and $z$-axis. In Section~\ref{sec:result_overlap}, we show the effects of varying the overlap parameters. If overlap is not sufficient, major artifacts start to appear in the reconstructions.

Given the size of each sub-volume and the amount of overlap desired, the number of parallel nodes ($M$) needed to cover the full volume is $M=M_xM_yM_z$, where
\begin{equation}
    M_x = \left\lceil\frac{1 - r_{vx}r_o}{r_{vx} - r_{vx}r_o}\right\rceil.
    \label{eq:M}
\end{equation}
$0 \leq r_{vx} \leq 1$ is the ratio of size of reconstructed volume to that of full volume in $x$, and $0 \leq r_o \leq 1$ is the overlap ratio between consecutive sub-volumes. $M_y$ and $M_z$ can be derived similarly. Notice that this is also the largest number of parallel nodes needed on the cluster to reconstruct the full volume. 

Another benefit that follows from being able to solve for sub-volumes is that one can easily reconstruct a custom volume-of-interest in order to quickly zoom in on particular features in 3D, or to avoid reconstructing areas that the sample does not occupy without requiring unnecessary computation. 
For instance, flat samples that have disproportionate ratio of length and width to depth can have reconstructions with larger $M_x, M_y$ and smaller $M_z$.

\subsection{Distributed algorithm}
We now describe the overall algorithm for our distributed reconstruction method and show the pseudo code in algorithm~\ref{alg:distributed_algorithm}. With sub-volumes having size $\{r_{vx}N_x, r_{vy}N_y, r_{vz}N_z\}$ and an overlap ratio $r_o$, we first calculate the centers of each sub-volume. As shown previously, along the $x$-axis, there are $M_x$ nodes, so for each node $m (1 \leq m \leq M_x)$, the center $x_m$ is:
\begin{equation}
    x_m = \frac{N_x}{2}\left(1 + (r_{vx}-r_or_{vx})(2m-M_x-1)\right).
    \label{eq:center}
\end{equation}
The centers are defined to be uniformly spaced. Similarly, volume centers along the $y$-axis and $z$-axis can be calculated as $y_l$ and $z_n$, respectively, where $1 \leq l \leq M_y$ and $1 \leq n \leq M_z$.

Next, we calculate the sub-volume center points' deviation from the center of the full volume, and apply Eq.~\eqref{eq:shfit_tilt} to obtain the cropped tilt-series corresponding to each sub-volume from the original tilt-series (see Fig.~\ref{Fig:schematics}). Each sub-volume may then be reconstructed independently of all other sub-volumes with any choice of tomographic reconstruction algorithm. As mentioned previously, we use the multi-slice algorithm~\cite{Ren:2020} to account for multiple scattering. After all sub-volumes are reconstructed, we stitch them together into the full volume reconstruction using a volume blending method described in~\cite{Chen:2020multi,Chowdhury:2019}.

\begin{algorithm}[htb]
    \caption{Distributed algorithm}
    \label{alg:distributed_algorithm}
    \textbf{Input:} Tilt angles $\{\theta_k\}_{k=1}^{N_{\theta}}$, measured intensity images $\{I_{k,j}\}_{k=1}^{N_{\theta}}$, sub-volume size $\{r_{vx}N_x, r_{vy}N_y, r_{vz}N_z\}$, overlap ratio $r_o$, reconstruction algorithm $\text{alg}_{\text{recon}}$, and stitching algorithm $\text{alg}_{\text{stitch}}$.
    \begin{algorithmic}[1] 
      \State $\left(\{x_m\}_{m=1}^{M_{x}},\{y_l\}_{l=1}^{M_{y}},\{z_n\}_{n=1}^{M_{z}}\right) \gets$  Eq.~\eqref{eq:center}\\
       \Comment{Pre-calculate centers}
      
      \For{each sub-volume center $i$}
          \State $x_0 \gets N_x/2 - x_m$
          \State $z_0 \gets N_z/2 - z_m$
          \For{$k\gets 1$ to $N_{\theta}$} \Comment{Shift and crop projections}
              \State $\Delta x \gets x_0\cos{\theta_k}+z_0\sin{\theta_k}$
              \State $\Delta y \gets y_m$
              \State $I_k' \gets \text{shift}\left(I_k, [\Delta x, \Delta y]\right)$
              \State $I_k' \gets \text{crop\_center}\left(I'_k, \text{size}=[r_{vx}N_x, r_{vy}N_y]\right)$
          \EndFor	
          \State $f_i^* \gets \text{alg}_{\text{recon}}\left(\{I'_k\}_{k=1}^{N_{\theta}}\right)$
    \EndFor	  
    \State $f^* \gets \text{alg}_{\text{stitch}}\left(\{f_i^*\}_{i=1}^{M}\right)$
    \end{algorithmic}
    \textbf{Return:} Estimated full volume $f^*$.
\end{algorithm}

\section{Results}
\begin{figure*}[!thbp]
	\centering
	\includegraphics[width=\textwidth]{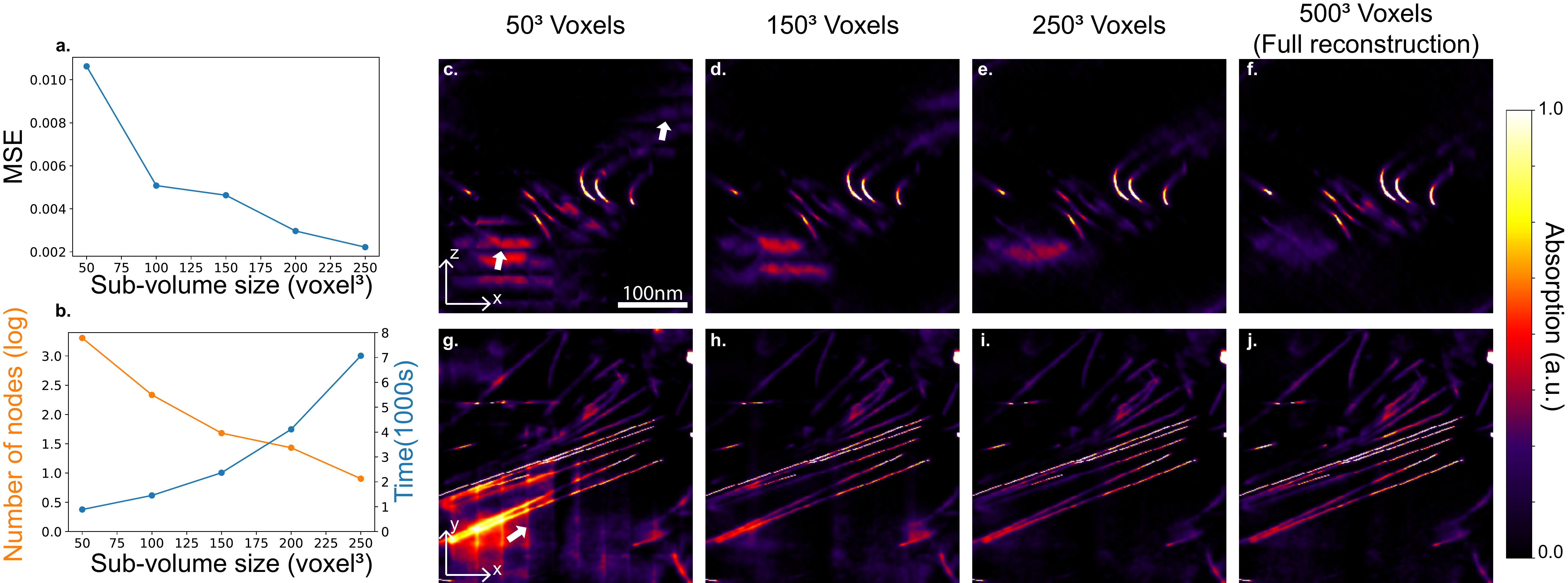}
	\caption{Splitting the volume into more sub-volumes, each with smaller size, enables faster reconstructions, but also induces more artifacts. (a) Extra reconstruction error (MSE with respect to a full reconstruction) vs. sub-volume size. (b) Number of nodes needed and reconstruction time for each node vs. sub-volume size. (c)-(f) Example $z-x$ slice and (g)-(j) $y-x$ slice of reconstructed 3D volume with various sub-volume sizes. As the sub-volume size decreases, artifacts increase, as indicated by the white arrows.}
	\label{Fig:result_size}
\end{figure*}

To test our distributed algorithm, we used the experimental data described in Whittaker et al.~\cite{Whittaker:2020}. The sample is clay minerals suspended in a vitrified solution of electrolyte. The sample is imaged using a Titan Krios TEM operated at 300 KeV. A Gatan K3 direct electron detector is used, which has an effective pixel size of 0.91 \AA~under superresolution mode. The images are then binned down by a factor of 2 after registration. After binning and removing the boundaries, each image has 5760$\times$4092 pixels. The full tilt series has 121 tilt angles and 3 defocus images per tilt, with a total applied dosage of 1100 $e^-$/\AA. The large field-of-view 3D reconstruction of the sample allows not only a direct comparison with the previously determined average static structure factor by \textit{in-situ} X-ray scattering, but also more insight into the dynamic mesoscale properties of the sample~\cite{Whittaker:2020}.


To reconstruct each sub-volume, we used the algorithm outlined in~\cite{Ren:2020} with a multi-slice forward model. Thus, it includes nonlinear modelling of the 3D sample that captures the multiple-scattering events between the probe and the sample. We then solve for the sample's 3D electric potential by formulating a nonconvex optimization algorithm. The multi-slice reconstruction algorithm is implemented using PyTorch~\cite{Pytorch:2019}, including GPU acceleration. The distributed algorithm, on the other hand, is implemented in Python using only CPU.

\begin{figure}[!htbp]
	\centering
	\includegraphics[width=0.7\linewidth]{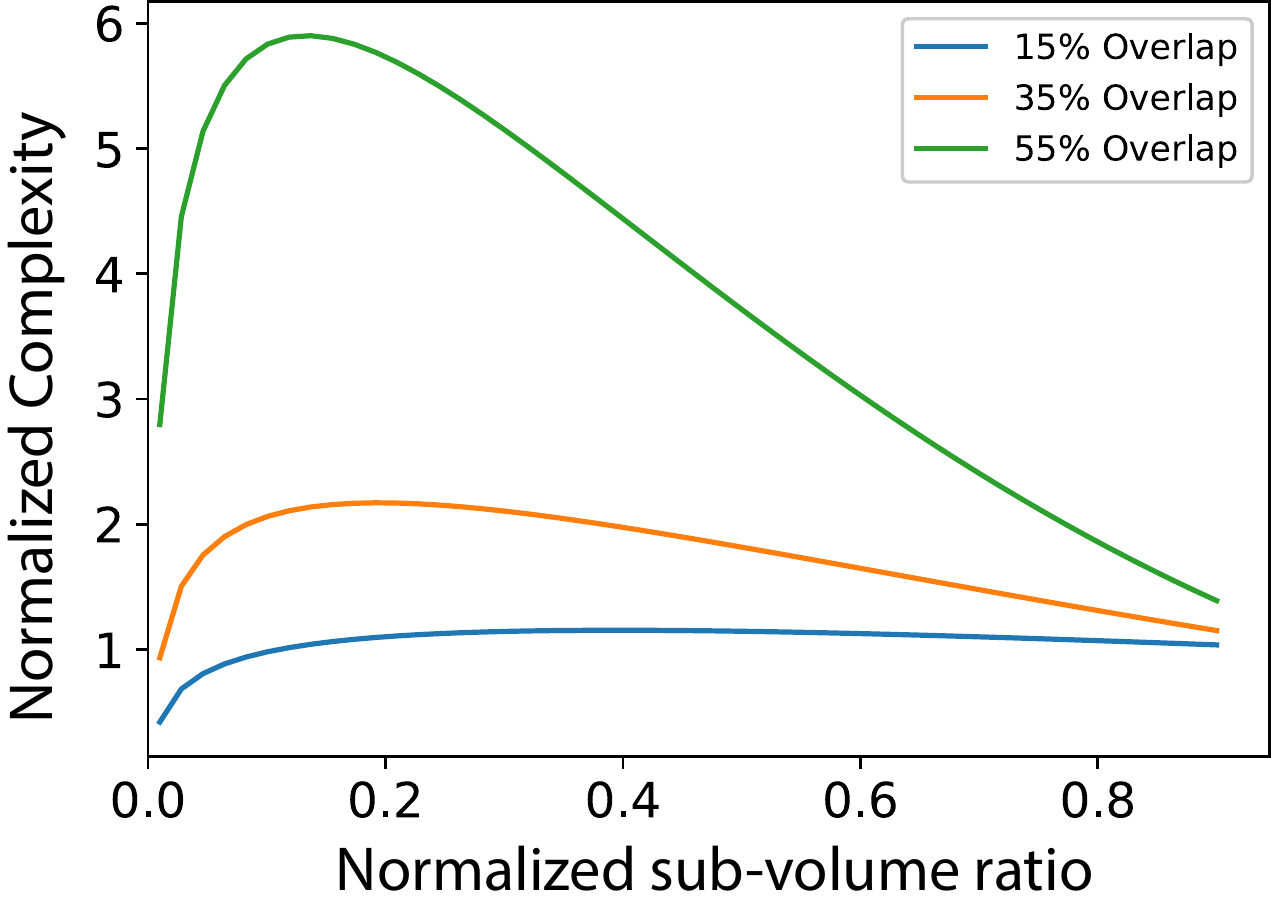}
	\caption{Computation complexity as a function of ratio of reconstructed sub-volume size to full volume size and sub-volume overlap ratio. The complexity is normalized with respect to that of a single reconstruction.}
	\label{Fig:result_complexity}
\end{figure}

The complexity of the multi-slice algorithm~\cite{Ren:2020} on a single volume is $\mathcal{O}\left(N\log{}N\right)$, where $N = N_xN_yN_z$ is the number of voxels in the reconstruction. By following the calculations in Section~\ref{subsec:overlap}, the total computation complexity across all compute nodes is therefore $\mathcal{O}\left(M(r_vN)\log{(r_vN)}\right)$, with sub-volume ratio $r_v = r_{vx}r_{vy}r_{vz}$ (ratio of the size of a sub-volume to that of the full volume) and $M$ nodes required to cover the full volume. Figure~\ref{Fig:result_complexity} illustrates the complexity of the proposed algorithm in comparison with the full tomographic reconstruction. For all overlap ratios ($r_o$) and sub-volume ratios ($r_v$) tested, our algorithm has higher complexity than if a single multi-slice reconstruction is run on the entire volume. However, since all $M$ tomographic reconstructions can be carried out simultaneously on different nodes of the compute cluster, the total run time for our algorithm is notably faster than for the full reconstruction.

All parallel reconstructions were performed using the Lawrencium high-performance computing facility at Lawrence Berkeley National Lab. The cluster contains nodes with one Intel 8-core Xeon Silver-4112 CPU and two NVIDIA V100 GPUs per node. The raw intensity measurements were first loaded into the RAM of a single node that then split it into multiple parallel job instances. Then, the jobs were distributed such that each individual node reconstructed one sub-volume. Finally, all results were aggregated on a single node to form the full volume using a linear stitching algorithm~\cite{Chen:2020multi,Chowdhury:2019}. 

\begin{figure*}[!thbp]
	\centering
	\includegraphics[width=\textwidth]{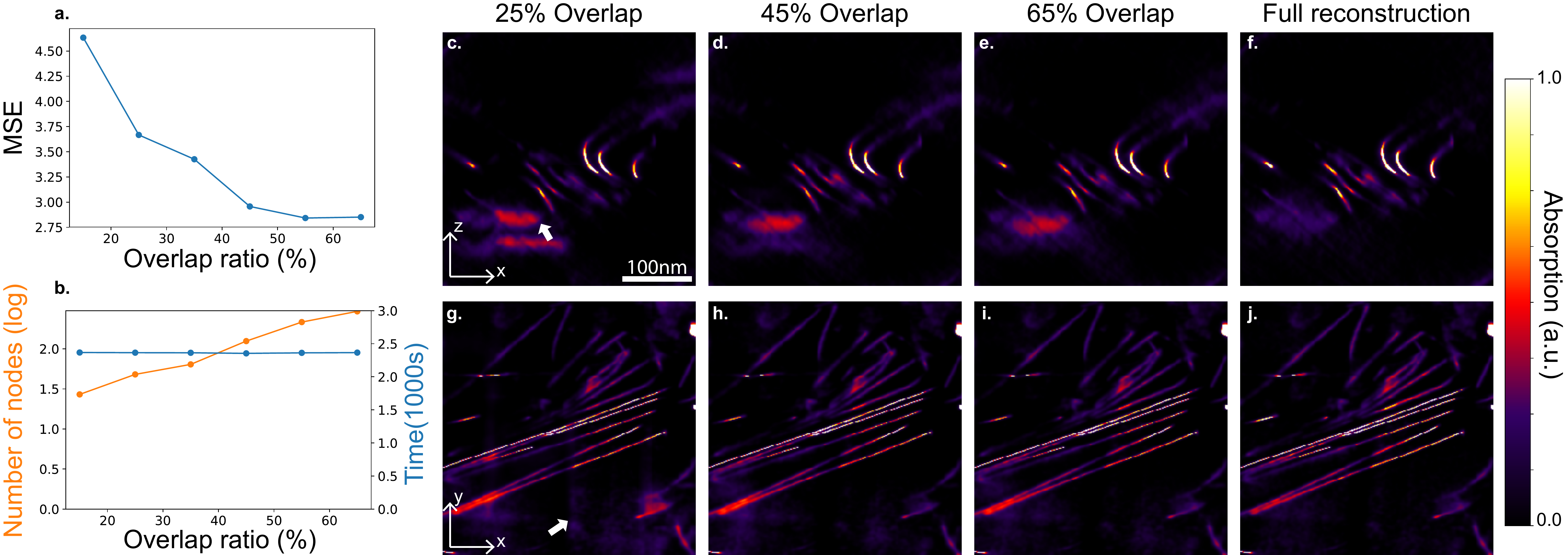}
	\caption{Increasing the overlap between adjacent sub-volumes reduces reconstruction errors. (a) Extra reconstruction error (MSE with respect to a single reconstruction) vs. overlap ratio. (b) Number of nodes needed and reconstruction time for each node as overlap ratio varies. (c)-(f) Single $z-x$ slice of reconstructed volume with various overlap ratios. (g)-(j) Single $y-x$ slice of reconstructed volume with various overlap ratios. (f) and (j) are full single reconstructions without any overlaps.}
	\label{Fig:result_overlap}
\end{figure*}

We tested the performance of the proposed algorithm by varying sub-volume size (Fig. \ref{Fig:result_size}) as well as overlap ratio (Fig. \ref{Fig:result_overlap}) and calculating the Mean Square Error (MSE) with respect to a single full tomographic reconstruction ($f^*$), which we treated as the ground truth in the error calculation:   
\begin{equation}
    MSE(f) = \frac{1}{N}\displaystyle\sum_{x,y,z}\left|f(x,y,z) - f^*(x,y,z)\right|^2.
\end{equation}
For these comparisons, the raw images were down-sampled by a factor of 4$\times$ in order to make a full tomographic reconstruction feasible on a single NVIDIA V100 GPU, and parallel reconstructions were carried out with only one reconstruction on a GPU at a time for benchmarking purposes. All reconstructions used 50 iterations to ensure convergence.

\subsection{Varying Sub-volume Size}
First, we compared the performance for different sizes of individual sub-volumes, $r_v$, which trade off parallelization (compute speed) for reconstruction quality. Specifically, we increased volume sizes from $50^3$ voxels to $250^3$ voxels, with the overlap ratio being fixed at 25\%. Figure~\ref{Fig:result_size} shows the result. As expected, decreasing the sub-volume size causes the reconstruction quality to deteriorate and artifacts to appear. This is because the proposed algorithm requires cropping of the projected image to reconstruct a reduced volume. However, electrons that scatter or diffract outside of the designated volume in 3D could still contribute to the projection data, and hence create artifacts during reconstruction. For larger sub-volume sizes, there is less contribution from outside of the sub-volume, and hence artifacts are reduced. 

As predicted by the complexity analysis, reconstruction time decreases when more nodes are used and the size of each sub-volume decreases (Fig.~\ref{Fig:result_size}(b)). Overall, many essential structures in the sample that appear in the full reconstruction are also present in the reconstructions at all sub-volume sizes. However, there are diminishing marginal returns on reconstruction time as we decrease the sub-volume size. When $N$ is sufficiently small, other operations in the algorithm start to dominate the reconstruction time, such as memory transfer, sample rotation, and regularization. The increasing severity of artifacts prevented us from reducing the size further.

\subsection{Varying Overlap Ratio}
\label{sec:result_overlap}
Next, we tested our algorithm's performance for varying overlap ratio, $r_o$. As mentioned in Section~\ref{subsec:overlap}, overlaps are necessary to take into account edge effects due to diffraction and multiple-scattering. The overlap ratio is varied from 15\% to 65\% while maintaining the sub-volume size to be a cube with $150^3$ voxels. Similar to the previous section, we compared the MSE and compute times when the parameter is changing, with results shown in Fig.~\ref{Fig:result_overlap}. As predicted, with only 25\% overlap (less than the minimum overlap of 30\% calculated earlier in Section~\ref{subsec:overlap}), the reconstruction suffers from some artifacts  (see white arrows in Fig.~\ref{Fig:result_overlap}(c) \& (g)). With overlap ratio greater or equal to 45\%, the MSE starts to converge. It is worth noting that the reconstruction time for individual sub-volume reconstructions are almost identical, because overlap ratio does not play a role in the complexity of individual sub-volume reconstructions; however, the number of nodes needed to cover the entire volume varies. Thus, the total complexity is still increasing as a function of overlap ratio. 

\subsection{Carbon Footprint}
\begin{figure}[!thbp]
	\centering
	\includegraphics[width=\linewidth]{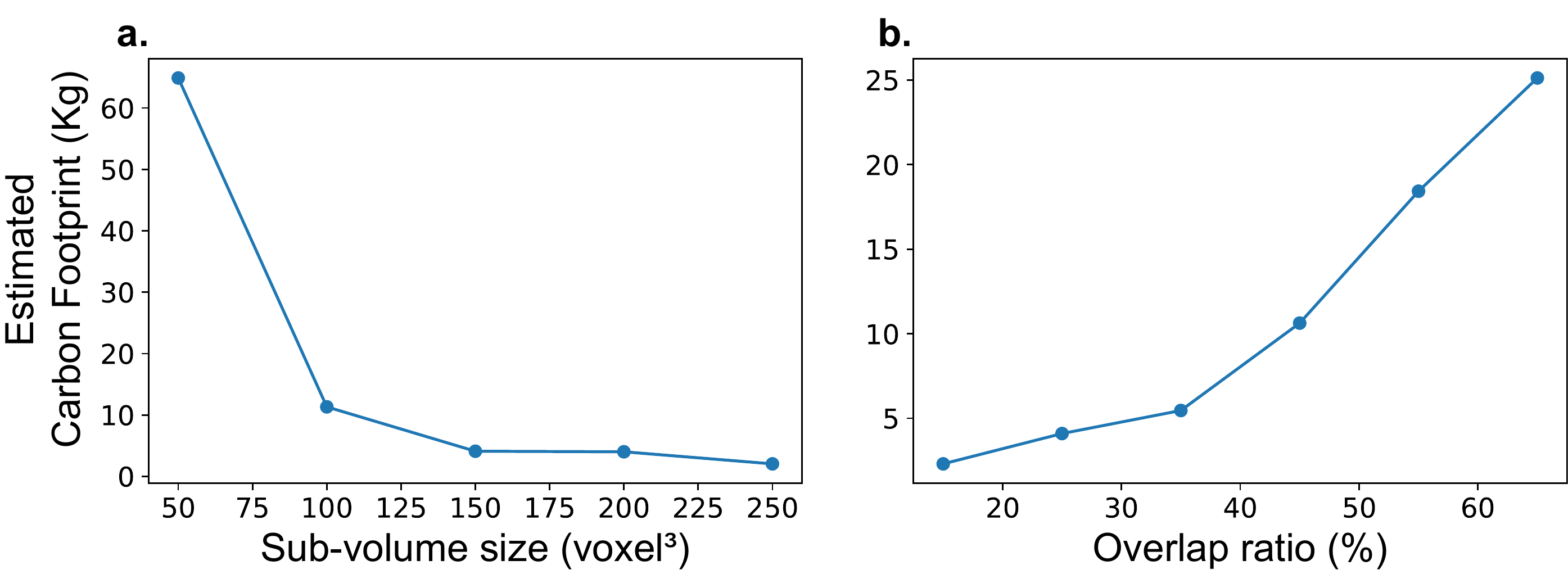}
	\caption{Carbon footprint with (a) varying sub-volume sizes and (b) varying overlap ratio. The optimal combination should be made by balancing the two parameters, such as sub-volume size of 150$^3$ voxels and a 35\% overlap ratio.}
	\label{Fig:result_carbon}
\end{figure}

Since our methods will use significant compute power for large-volume high-resolution datasets, we report the equivalent carbon dioxide (CO$_2$) emission of the previous experiments. All numbers are calculated in units of Kilogram (Kg), and as a baseline we assume that each NVIDIA GPU V100 has a carbon footprint of 0.13 Kg/hr. Estimations were conducted using the \href{https://mlco2.github.io/impact#compute}{MachineLearning Impact calculator} presented in \cite{Lacoste:2019carbon}. The total CO$_2$ emission is the baseline multiplied by total amount of computation taken across all nodes, as shown in Fig.~\ref{Fig:result_carbon}. Figure~\ref{Fig:result_carbon}(a) shows the carbon footprint for experiments that vary sub-volume sizes, and (b) shows the experiments that vary overlap ratio. In Fig.~\ref{Fig:result_carbon}(a), the carbon dioxide emission converges above $100^3$ voxels, because the increase in number of nodes $M$ trades off with reduced time needed to reconstruct smaller volumes. However, the trade-off is no longer true for sizes smaller than $100^3$ voxels, as computation overhead starts to dominate. The curve in Fig.~\ref{Fig:result_carbon}(b) mostly adheres to the predicted result - since the size of sub-volumes remain the same for any overlap ratio, the CO$_2$ emission is scaling linearly with the number of nodes in order to cover the full volume. Therefore, the overlap ratio should be chosen to be as small as possible, yet satisfying the minimum derived in Section~~\ref{subsec:overlap} to achieve an acceptable reconstruction quality.

\begin{figure*}[!thbp]
	\centering
	\includegraphics[width=\textwidth]{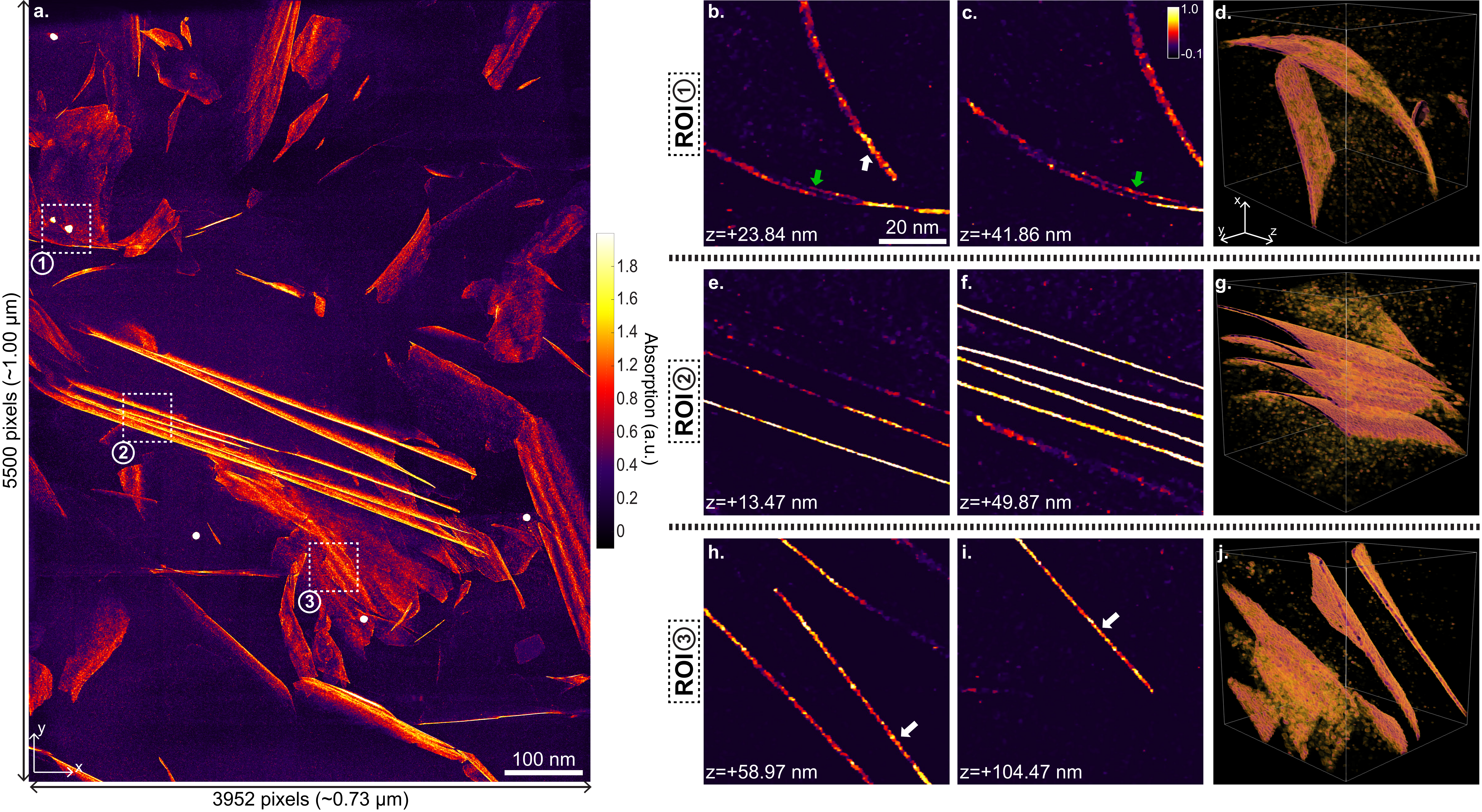}
	\caption{Full field-of-view 3D reconstruction of lithium-montmorillonite, with a total of 5500$\times$3952$\times$952 voxels. (a) Maximum depth projection of the absorption channel. Three smaller regions-of-interest (ROIs) are cropped and zoomed in, with sizes of $400^3$ voxels. (b)-(c),(e)-(f),and (h)-(i) Show zoom-in lateral slices at different depth for ROI 1, 2, and 3, respectively. (d), (g), and (j) show 3D volume renders for each of the ROIs.}
	\label{Fig:result_large_recon}
\end{figure*}
\subsection{Full Resolution Reconstruction}
We show in Fig.~\ref{Fig:result_large_recon} a reconstruction of the full lithium-montmorillonite dataset in a volume of $3952\times5500\times952$ voxels with isotropic resolution of $1.82$ $\AA^3$/voxel. This corresponds to a volume of $0.73(x)\times1.00(y)\times1.73(z) \mu \text{m}^3$. Figure~\ref{Fig:result_large_recon}(a) shows the maximum projection of the volume along the $z$-axis. All features are reconstructed, with occasional artifacts observed in the background, either due to noise or the stitching algorithm. We zoom in on three regions-of-interest (ROIs) in Fig.~\ref{Fig:result_large_recon}(b-j). Within each, the clay layers have different degrees of curvature and number of neighboring layers. For each sub-volume, two cross-sections at different depths are shown, along with a volume render. As pointed out with green arrows in Fig.~\ref{Fig:result_large_recon}(b,c), two layers that are closely stacked ($\sim$ 1nm separation) can be observed in ROI 1. If the full volume was reconstructed using the downsampled tilt-series, this observation would not be possible. Also enabled by high-resolution reconstruction, detailed crystal structures that were previously blurred out in~\cite{Whittaker:2020} are now visible, as indicated by white arrows in Fig.\ref{Fig:result_large_recon}(b,h and i).

Notice that the total depth of the sample is significantly smaller than the other two dimensions, since the sample is fairly flat. Because our method is capable of reconstructing a custom-sized volume, a smaller depth is chosen such that computation resources are not wasted on empty space. The volume is broken down into 702 nodes of cubic volume reconstructions, with an isotropic overlap of 25\%. Each sub-volume has a size of $500^3$ voxels. The full volume stored in complex 32 bit float has a size of $\sim$154 GB, which is difficult to fit into the RAM of a computer, particularly with the overhead storage requirements of the algorithm. In addition, GPU acceleration is often needed in order for the algorithms to converge in a reasonable amount of time, and volumes with such sizes cannot be fit into the memory of a modern GPU without partitioning.

\section{Discussion}
The distributed computing algorithm we proposed allows large tomography datasets to be decomposed into smaller independent sub-problems for faster compute times on clusters. One can choose different values of overlap ratio ($r_o$) and sub-volume sizes ($r_v$) according to their tolerance of artifacts, or optimize parameters to minimize the overall carbon footprint while ensuring a reasonable reconstruction quality. 

Once the full tilt-series is split into multiple independent projection data, reconstructions are carried out in parallel by any tomography algorithm of choice. In our work, to accurately model the interaction between the electron beam and the clay minerals in the large volume at microscopic level, we adopted the multi-slice algorithm~\cite{Ren:2020} that is capable of accounting for multiple-scattering samples. Our distributed algorithm is similar to that in X-ray CT by Basu and Bresler~\cite{Basu:2000n,Basu:2002n}; however, we use coherent multiple-scattering reconstruction algorithms instead of the filtered back-projection algorithm. We also do not continue to decompose the tomography problem in a hierarchical manner for further parallelization, as in ~\cite{Basu:2000n,Basu:2002n} because we find that there is a limit as to how small a sub-volume can be without suffering from severe artifacts. Further acceleration could be achieved by combining the slice-binning idea outlined in~\cite{Ren:2020}. In this method, consecutive axial slices at each tilt angle are summed into a ``thicker" slice during the forward propagation. In the update step, the error gradient is equally distributed back to individual slices. Slice-binning allows significant reduction of number of slices required to propagate through the 3D sample, without significant loss of accuracy.

To reduce the reconstruction artifacts, effort can be spent on exploring more advanced algorithms for volume stitching. In our study, we used a simple weighting function for all sub-volumes similar to~\cite{Chen:2020multi,Chowdhury:2019}. More complicated algorithms such as pyramid blending, or 2-band blending, could be explored~\cite{Gracias:2009fast,Efros:2001image}. However, the major drawback of these methods is that they are content-aware to some extent, and could be reducing reconstruction accuracy in order to reduce stitching artifacts near the boundary. 

\section{Conclusion}
In this work, we described a distributed computing strategy for 3D reconstruction from intensity-only TEM tilt-series data. We perform projection data manipulation as a pre-processing step before reconstructing many sub-volumes simultaneously on a compute cluster. In the end, all sub-volumes are stitched together to form the full reconstruction. By combining this method with a multiple-scattering reconstruction algorithm, we have successfully demonstrated the validity of the algorithm. We compared various parameters on experimental data and showed performance difference in terms of MSE and reconstruction time, and we reconstructed a large 3D volume of size  ($0.73(x)\times1.00(y)\times1.73(z) \mu \text{m}^3$) with resolution of 1.82 $\rm{\AA}^3$/voxel, the largest cryo-ET reconstruction to our knowledge. With minimal restrictions for the scattering algorithm of choice, and minimal knowledge required of computer architecture, this method opens the door to large tomographic reconstructions that require heavy computational resources, and provides great flexibility in choosing specific volumes-of-interest to recover.



\section*{Acknowledgment}
This work was supported by STROBE: A National Science Foundation Science \& Technology Center under Grant No. DMR 1548924; U.S. Department of Energy, Office of Science, Office of Basic Energy Sciences, Chemical Sciences, Geosciences, and Biosciences Division, through its Geoscience program at LBNL under Contract DE-AC02-05CH11231. Work at the Molecular Foundry was supported by the Office of Science, Office of Basic Energy Sciences, of the U.S. Department of Energy under Contract No. DE-AC02-05CH11231. We also wish to thank the NVIDIA Corporation for donation of GPU resources.

\ifCLASSOPTIONcaptionsoff
  \newpage
\fi



\bibliographystyle{IEEEtran}
\bibliography{ref}
%



%
\newpage
\begin{IEEEbiography}[{\includegraphics[width=1in,height=1.25in,clip,keepaspectratio]{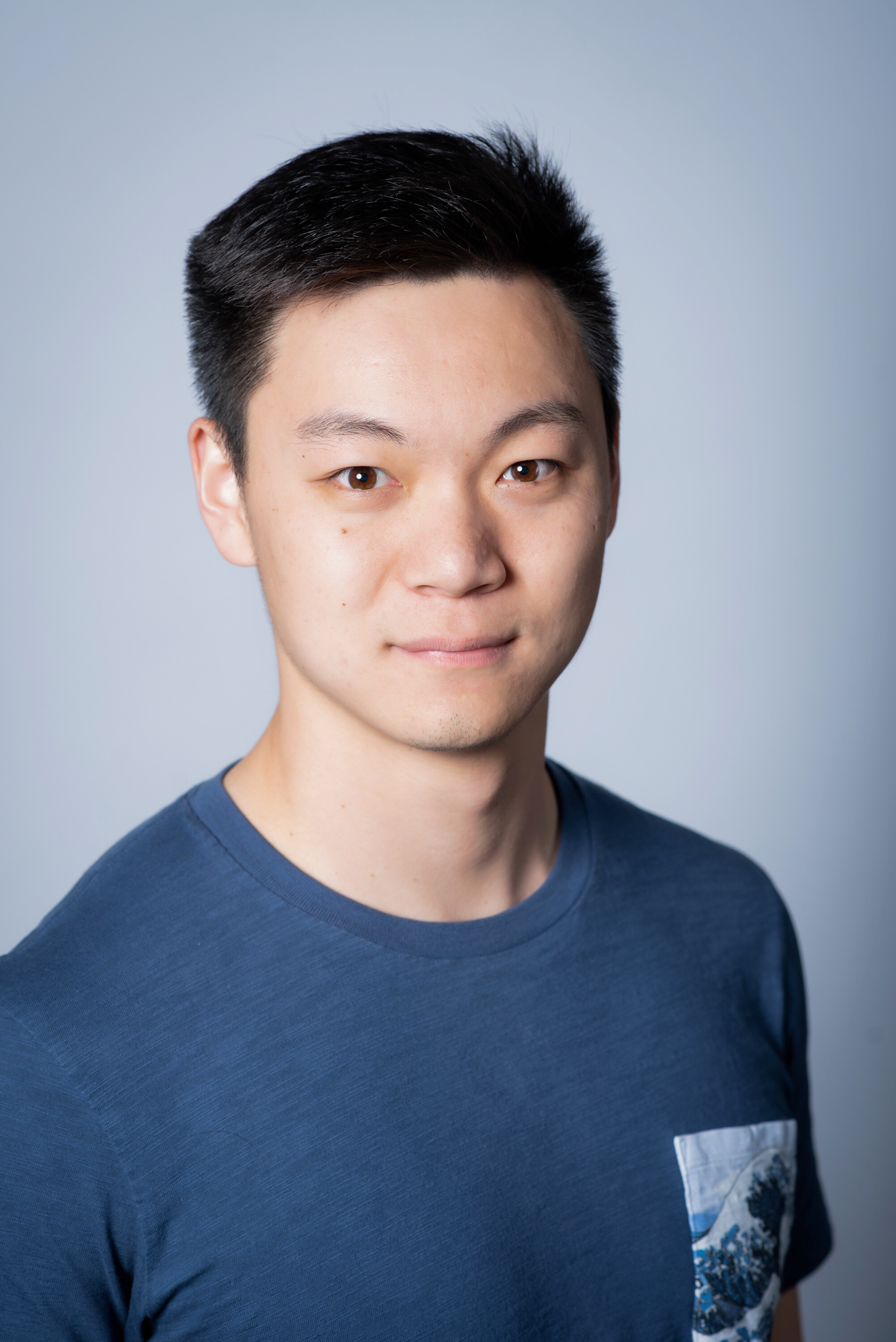}}]{David Ren}
is a graduate researcher in the deparment of Electrical Engineering and Computer Sciences at UC Berkeley, and affiliated with the National Center for Electron Microscopy, Molecular Foundry at Lawrence Berkeley National Laboratory (LBNL). He received his B.S. degrees from the University of Illinois at Urbana-Champaign in Electrical \& Computer Engineering and Applied Mathematics in 2016. His research focuses on developing algorithms to achieve high throughput 3D electron tomography for multiple scattering samples.
\end{IEEEbiography}

\begin{IEEEbiography}[{\includegraphics[width=1in,height=1.25in,clip,keepaspectratio]{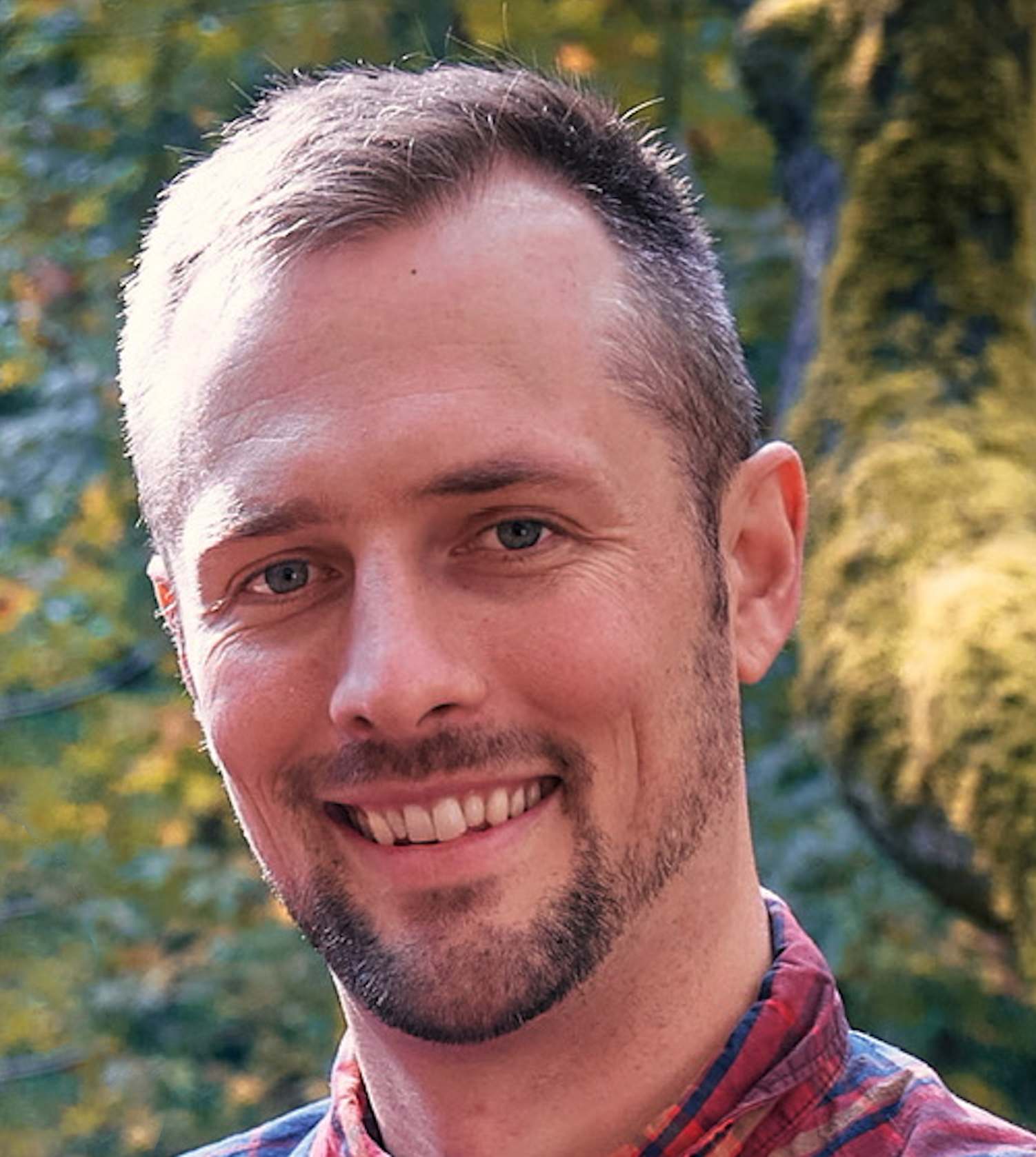}}]{Michael L. Whittaker}
is a research scientist in the Energy Geoscience Division (EGD) in the Earth and Environmental Sciences Area (EESA) at Lawrence Berkeley National Laboratory (LBL) and an affiliate in the department of Earth and Planetary Science (EPS) at the University of California, Berkeley (UCB). He received BS/MS degrees from the University of Utah in Materials Science and Engineering in 2012, and Ph.D. from Northwestern University in Materials Science and Engineering in 2017 with advisor Derk Joester. Mike worked as a postdoc in the nanogeoscience group in EGD from 2017-2020 with Benjamin Gilbert and with Jill Banfield at UCB. He leads the geomaterials research group is the cofounder of LiRRIC, the Lithium Resource Research and Innovation Center at LBL..
\end{IEEEbiography}

\begin{IEEEbiography}[{\includegraphics[width=1in,height=1.25in,clip,keepaspectratio]{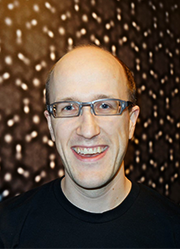}}]{Colin Ophus}
received his PhD in Materials Engineering from the University of Alberta in Canada. He is currently a staff scientist at the National Center for Electron Microscopy, part of the Molecular Foundry at Lawrence Berkeley National Laboratory. He primarily works on developing methods, algorithms, and codes for simulation, analysis, and instrument design for high resolution and scanning transmission electron microscopy.

\end{IEEEbiography}


\begin{IEEEbiography}[{\includegraphics[width=1in,height=1.25in,clip,keepaspectratio]{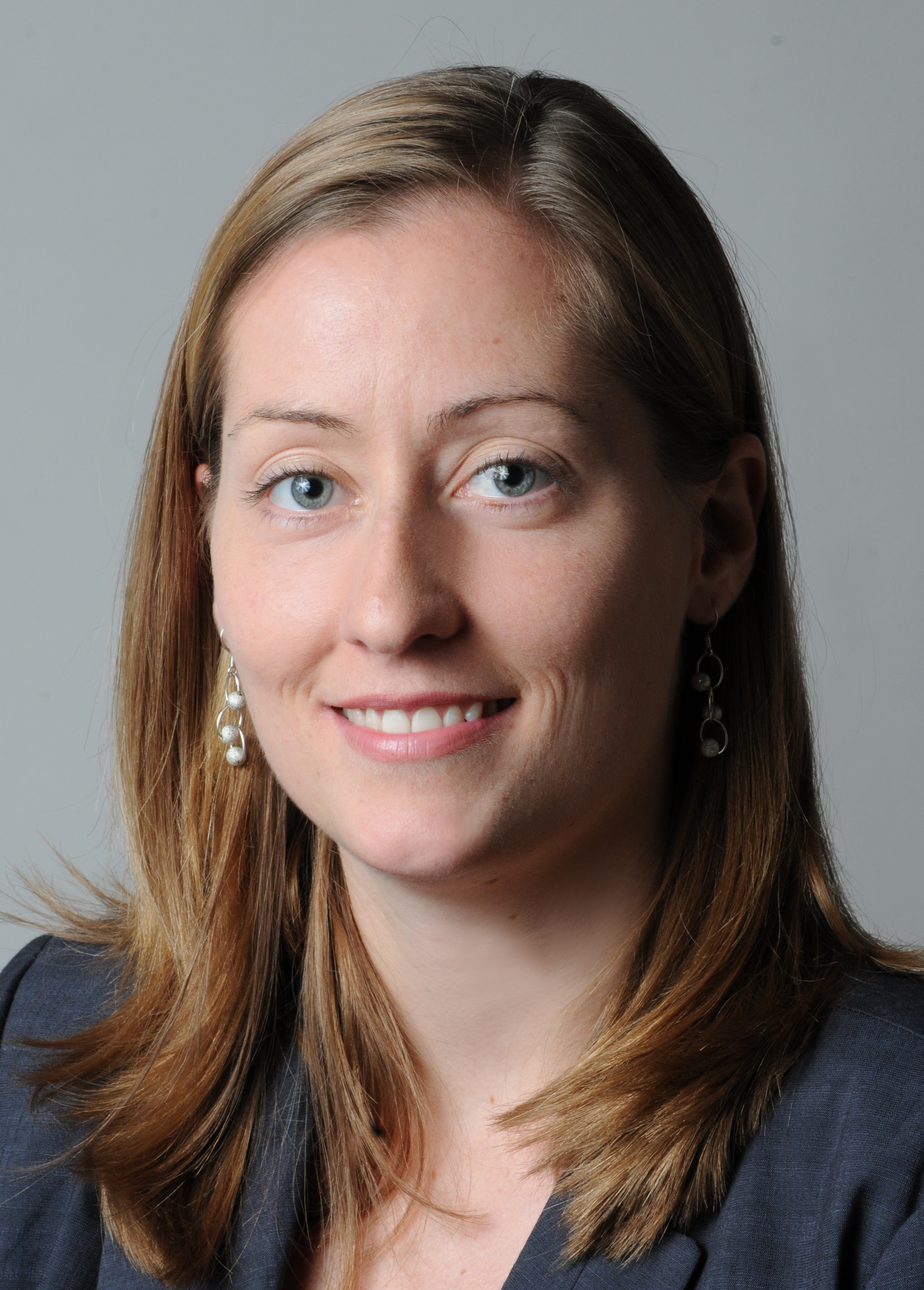}}]{Laura Waller}
is an Associate Professor of Electrical Engineering and Computer Sciences (EECS) at UC Berkeley, and affiliated with the UCB/UCSF Bioengineering Graduate Group and Applied Sciences \& Technology program. She received B.S., M.Eng. and Ph.D. degrees from the Massachusetts Institute of Technology (MIT) in 2004, 2005 and 2010, and was a Postdoctoral Researcher and Lecturer of Physics at Princeton University from 2010-2012. She is a Packard Fellow for Science \& Engineering, Moore Foundation Data-driven Investigator, Bakar Fellow, OSA Fellow, AIMBE Fellow and Chan-Zuckerberg Biohub Investigator. She has recieved the Carol D. Soc Distinguished Graduate Mentoring Award, Agilent Early Career Profeessor Award (Finalist), OSA Adolph Lomb Medal, Ted Van Duzer Endowed Professorship, NSF CAREER Award and the SPIE Early Career Achievement Award. 
\end{IEEEbiography}




\end{document}